\documentclass[12pt]{article}

\usepackage{physics}
\usepackage[utf8]{inputenc}
\usepackage[T1]{fontenc}
\usepackage{cite}
\usepackage{graphicx}
\usepackage{array}
\usepackage{float}
\newcommand\Tstrut{\rule{0pt}{2.6ex}}
\newcommand\Bstrut{\rule[-0.9ex]{0pt}{0pt}} 
\newcommand\Tstrutt{\rule{0pt}{3.1ex}}
\newcommand\Bstrutt{\rule[-1.4ex]{0pt}{0pt}} 
\newcommand*{\sk}{-3pt}

\title{\vspace*{-40pt}\textbf{Quark unitarity triangles}} \date{}
\author{S.~Rebeca Juárez Wysozka\\
  \small{Departamento de Física, Escuela Superior de Física y
    Matemáticas}\\[\sk]
  \small{Instituto Politécnico Nacional. U.P Adolfo López Mateos}\\[\sk]
  \small{C.P. 07738. Ciudad de México, México}\\[3pt]
  Piotr Kielanowski\\
  \small{Departamento de Física, Centro de Investigación y
    Estudios Avanzados}\\[\sk]
  \small{Av. Instituto Politécnico Nacional 2508, C.P. 07000}\\[\sk]
  \small{Ciudad de México, México}\\[3pt]
  Liliana Vázquez Mercado\\[\sk]
  \small{Departamento de Física, Centro Universitario de Ciencias
    Exactas e Ingenierías}\\[\sk]
  \small{Universidad de Guadalajara, Av. Revolución 1500, Colonia
    Olímpica, C.P. 44430}\\[\sk]
  \small{Guadalajara, Jalisco, Mexico}\\
  \small{\textit{E-mail:} \texttt{rebecajw@gmail.com,
    piotr.kielanowski@cinvestav.mx,}}\\[\sk]
  \small{\texttt{liliana.vmercado@academicos.udg.mx}}}

\begin{document}
\maketitle
\begin{abstract}
  The angles of all unitarity triangles of the
  Cabibbo-Kobayashi-Maskawa matrix are determined from the
  experimental data. Our analysis is independent of the
  parameterization of the CKM matrix and it is based on the
  predictions of the unitarity for the angles and the areas of the
  unitarity triangles. We note that the lengths of the sides of the
  four unitarity triangles determined from the experimental data do
  not form a triangle. We resolve this incompatibility by performing a
  constrained fit, assuming the equality of the area of the unitarity
  triangles. We demonstrate that the measured data are compatible with
  the predictions of the unitarity of the Cabibbo-Kobayashi-Maskawa
  matrix, but there is a $2\sigma$ tension for one of the
  triangles. We show that the angles of the unitarity triangles
  obtained by the multiplication of the rows of the CKM matrix can be
  obtained from the angles obtained by the multiplication of the
  columns. The equality of those two types of the angles is a simple,
  but a very powerful test of the general structure of the Standard
  Model.
\end{abstract}

\section{Introduction}
\label{sec:introduction}

One of the most important challenges of the contemporary particle
physics is the search for the phenomena that cannot be explained by
the Standard Model~(SM). The main reason of such a search is that from
the theoretical standpoint the SM cannot be the final theory,
because it cannot explain many existing phenomena, like massive
neutrinos or dark matter. On the other hand there does not exist any
clear cut experimental result that contradicts the predictions of
the~SM.  A discovery of a contradictory result would serve a double
purpose. First, it would be a convincing proof of an inadequacy of the
SM for the description of all the elementary particles
phenomena. Second, it would be a clue on which possible extension of
the~SM to choose.

The structure of the paper is the
following. Section~\ref{sec:cabibbo-kobay-mask} contains the
introductory material in which we define the notation and discuss the
Cabibbo-Kobayashi-Maskawa (CKM) matrix. In
Section~\ref{sec:unitarity-triangles} we consider the unitarity
triangles from the theoretical viewpoint and review the existing
experimental data for the CKM matrix. We also determine the lengths of
the sides of all unitarity triangles, using as an input the
experimental values of the absolute values of the CKM matrix elements.
Section~\ref{sec:analys-triangle-delt} contains the comparison of two
unitarity triangles $\Delta_{2}$ and $\Delta_{5}$. The triangle
$\Delta_{2}$ is the only one, whose angles have been experimentally
measured and from the unitarity of the CKM matrix it follows that the
angles $\phi_{2}$ of the $\Delta_{2}$ and $\Delta_{5}$ triangles
should be equal. We also determine the lengths of the sides of the
triangle $\Delta_{2}$, but this time, using as input the angles and
the lengths of the triangle. In
Section~\ref{sec:analysis-remaining-triangles} we discuss the
properties of the remaining unitarity
triangles. Section~\ref{sec:disc-concl} contains the discussion of
results and final remarks.

\section{Cabibbo-Kobayashi-Maskawa matrix}
\label{sec:cabibbo-kobay-mask}

The interactions of quarks with charged vector bosons $W_{\mu}^{\pm}$
are described in the SM~\cite{bib:2, bib:3, bib:4, bib:5, bib:6,
  bib:7, bib:8} by the Cabibbo-Kobayashi-Maskawa
(CKM)~\cite{bib:9,bib:9a,bib:1,bib:1a,bib:34a} matrix $V_{\text{CKM}}$
\begin{equation}
  \label{eq:1}
  V_{\text{CKM}}=
  \mqty(V_{ud}&V_{us}&V_{ub}\\
  V_{cd}&V_{cs}&V_{cb}\\
  V_{td}&V_{ts}&V_{tb}),
\end{equation}
which is obtained from the quark Yukawa couplings by the bi-unitary
transformation. The CKM matrix is by construction a $3\times3$ unitary
matrix. The unitarity of a matrix means that the rows and columns are
normalized to~$1$ and are mutually orthogonal.  The verification of
the unitarity of the CKM matrix is an important tests of the SM model
and it consists in checking the orthonormality of the rows and
columns:
\begin{enumerate}
\item The length of each row and column should be~1. If it is not
  equal to~1 we have two possibilities
  \begin{enumerate}
  \item If it exceeds 1, then the universality of weak interactions
    between the quarks and leptons is violated.
  \item If it is less than 1, then it may be a sign of the existence
    of more than 3 generations of quarks or it may also be a sign of
    the universality violation.
  \end{enumerate}
\item The orthogonality of the rows and columns of the CKM matrix
  leads to the \textit{unitarity triangles}, i.e., the sum of the
  products of the CKM matrix elements of two different rows or columns
  have to be equal to~$0$. E.g., if we multiply the first column by
  the complex conjugate of the third column then we obtain
  \begin{equation}
    \label{eq:2}
    V_{ud}^{\phantom{*}}V_{ub}^{*}+V_{cd}^{\phantom{*}}V_{cb}^{*}
    +V_{td}^{\phantom{*}}V_{tb}^{*}=0,
  \end{equation}
  which on the complex plane is graphically represented as a triangle
  in Fig.~\ref{fig:1}.
  \begin{figure}[H]\centering
    \includegraphics[width=0.6\linewidth]{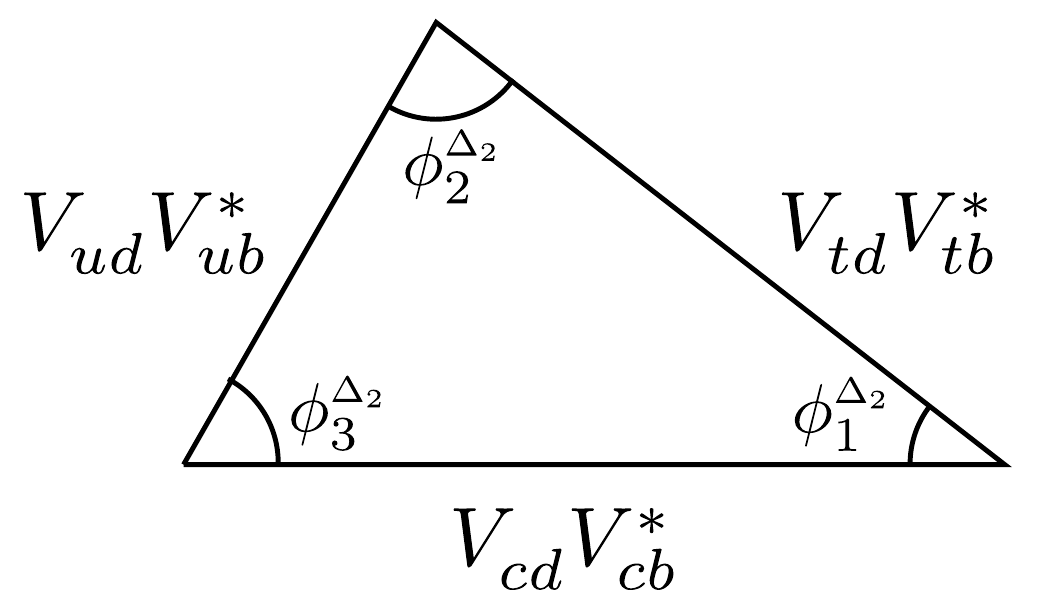}
    \caption{\label{fig:1}Unitarity triangle $\Delta_{2}$ (see
     Eq.~\eqref{eq:3a}). For other unitarity triangles, the sides are
     denoted according to Eqs.~\eqref{eq:3} and the angles carry the
     corresponding index~$\Delta_{i}$.}
  \end{figure}
\end{enumerate}

\section{Unitarity triangles}
\label{sec:unitarity-triangles}

For the CKM matrix one can construct 6 unitarity triangles $\Delta_{i}$:\\
A. \textit{Obtained from the columns multiplication}
\begin{subequations}\label{eq:3}
  \begin{equation}
    \begin{aligned}
     &\text{triangle }\Delta_{1}\qq*{:}
     V_{ud}^{\phantom{*}}V_{us}^{*}+V_{cd}^{\phantom{*}}V_{cs}^{*}
     +V_{td}^{\phantom{*}}V_{ts}^{*}=0, \\
     &\text{triangle }\Delta_{2}\qq*{:}
     V_{ud}^{\phantom{*}}V_{ub}^{*}+V_{cd}^{\phantom{*}}V_{cb}^{*}
     +V_{td}^{\phantom{*}}V_{tb}^{*}=0,\\
     &\text{triangle }\Delta_{3}\qq*{:}
     V_{us}^{\phantom{*}}V_{ub}^{*}+V_{cs}^{\phantom{*}}V_{cb}^{*}
     +V_{ts}^{\phantom{*}}V_{tb}^{*}=0,
    \end{aligned}\label{eq:3a}
  \end{equation}
  B. \textit{Obtained from the rows multiplication}
  \begin{equation}
    \begin{aligned}
     &\text{triangle }\Delta_{4}\qq*{:}
     V_{ud}^{\phantom{*}}V_{cd}^{*}+V_{us}^{\phantom{*}}V_{cs}^{*}
     +V_{ub}^{\phantom{*}}V_{cb}^{*}=0, \\
     &\text{triangle }\Delta_{5}\qq*{:}
     V_{ud}^{\phantom{*}}V_{td}^{*}+V_{us}^{\phantom{*}}V_{ts}^{*}
     +V_{ub}^{\phantom{*}}V_{tb}^{*}=0,\\
     &\text{triangle }\Delta_{6}\qq*{:}
     V_{cd}^{\phantom{*}}V_{td}^{*}+V_{cs}^{\phantom{*}}V_{ts}^{*}
     +V_{cb}^{\phantom{*}}V_{tb}^{*}=0.
    \end{aligned}\label{eq:3b}
  \end{equation}
\end{subequations}
From the unitarity of the CKM matrix it follows that all the unitarity
triangles have the same area which is equal to $J/2$, where $J$ is the
Jarlskog invariant\cite{bib:10, bib:11}.

The angles of the unitarity triangles are not independent and there
are simple relations that follow directly from the construction of the
unitarity triangles. Let us consider as an example the angles
$\phi_{1}^{\Delta_{1}}$ and $\phi_{3}^{\Delta_{6}}$
\begin{subequations}\label{eq:17}
  \begin{align}
    \label{eq:17a}
    \phi_{1}^{\Delta_{1}}
    &=-\arg\left(\frac{V_{td}^{\phantom{*}}V_{ts}^{*}} {V_{cd}^{\phantom{*}}V_{cs}^{*}}\right)
     = -\arg\left(V_{td}^{\phantom{*}}V_{ts}^{*}V_{cd}^{*}V_{cs}^{\phantom{*}}\right),\\[3pt]
    \label{eq:17b}
    \phi_{3}^{\Delta_{6}}
    &=-\arg\left(\frac{V_{cs}^{\phantom{*}}V_{ts}^{*}} {V_{cd}^{\phantom{*}}V_{td}^{*}}\right)
     = -\arg\left(V_{cs}^{\phantom{*}}V_{ts}^{*}V_{cd}^{*}V_{td}^{\phantom{*}}\right)
  \end{align}
\end{subequations}
and we see that these two angles are equal. In the same
way~\cite{bib:34,bib:35} we derive the following set of the relations
\begin{equation}
  \label{eq:14}
  \begin{aligned}
    \phi_{1}^{\Delta_{4}}&=\phi_{3}^{\Delta_{3}},\quad&
    \phi_{2}^{\Delta_{4}}&=\phi_{3}^{\Delta_{2}},\quad&
    \phi_{3}^{\Delta_{4}}&=\phi_{3}^{\Delta_{1}},\\
    \phi_{1}^{\Delta_{5}}&=\phi_{2}^{\Delta_{3}},&
    \phi_{2}^{\Delta_{5}}&=\phi_{2}^{\Delta_{2}},&
    \phi_{3}^{\Delta_{5}}&=\phi_{2}^{\Delta_{1}}\\
    \phi_{1}^{\Delta_{6}}&=\phi_{1}^{\Delta_{3}},&
    \phi_{2}^{\Delta_{6}}&=\phi_{1}^{\Delta_{2}},&
    \phi_{3}^{\Delta_{6}}&=\phi_{1}^{\Delta_{1}}.
  \end{aligned}
\end{equation}
Thus the angles of the unitarity triangles $\Delta_{4}$, $\Delta_{5}$,
$\Delta_{6}$, derived by the multiplication of the rows, can be
obtained from the angles of the triangles $\Delta_{1}$, $\Delta_{2}$,
$\Delta_{3}$, derived by the multiplication of the columns of the CKM
matrix.

The angles of the unitarity triangles were measured only for the
triangle $\Delta_{2}$ and their experimental values
$\phi_{i}^{\text{exp}}$ are~\cite{bib:1}:
\begin{equation}
  \label{eq:8}
  \sin(2\phi_{1}^{\text{exp}})=0.699\pm 0.017, \quad
  \phi_{2}^{\text{exp}}=(84.9\pm 5.1)^{\circ}\quad \phi_{3}^{\text{exp}}=(72.1\pm 4.5)^{\circ}
\end{equation}
and the lengths of the sides of all the triangles $\Delta_{i}$ can be
calculated from the experimentally known absolute values of the CKM
matrix elements which are equal to~\cite{bib:1}:
\begin{equation}
  \label{eq:6}
  \begin{aligned}
    \abs{V_{ud}^{\text{exp}}}&=0.97370\pm 0.00014,\quad
    &&\abs{V_{us}^{\text{exp}}}=0.2245\pm 0.0008,\\
    \abs{V_{ub}^{\text{exp}}}&=(3.82\pm 0.24)\times10^{-3},
    &&\abs{V_{cd}^{\text{exp}}}=0.221\pm0.004,\\
    \abs{V_{cs}^{\text{exp}}}&=0.987\pm0.012,
    &&\abs{V_{cb}^{\text{exp}}}=(41.0\pm 1.4)\times10^{-3},\\
    \abs{V_{td}^{\text{exp}}}&=(8.0\pm0.3)\times10^{-3},
    &&\abs{V_{ts}^{\text{exp}}}=(38.8\pm 1.1)\times10^{-3},\\
    \abs{V_{tb}^{\text{exp}}}&=1.013\pm0.03
  \end{aligned}
\end{equation}
From the values in Eq.~\eqref{eq:6} we obtain the lengths of the sides
of the unitarity triangles, given in Table~\ref{tab:1}.
\begin{table}[H]\centering
  \begin{tabular}{cccc}\hline
    Triangle\Tstrut\Bstrut&$l_{1}$&$l_{2}$&$l_{3}$\\
    \hline
    $\Delta_{1}$
    \Tstrut&$0.21860\pm0.00078$&$0.2181\pm0.0046$&$(3.10\pm0.15) \times 10^{-4}$\\
    $\Delta_{2}$&$(3.72\pm 0.23)\times 10^{-3}$ &$(9.06 \pm 0.35)\times 10^{-3}$&$(8.10 \pm 0.39)\times 10^{-3}$\\
    $\Delta_{3}$&$(8.58\pm0.54) \times 10^{-4}$&$0.0405\pm 0.0015$&$0.0393\pm 0.0011$\\
    $\Delta_{4}$&$0.2152\pm 0.0039$&$0.2216\pm 0.0026$&$(1.56\pm 0.11)\times 10^{-4}$\\
    $\Delta_{5}$&$(7.79\pm 0.29)\times 10^{-3}$&$(8.71\pm 0.25)\times 10^{-3}$&$(3.87\pm 0.27)\times 10^{-3}$\\
    $\Delta_{6}$\Bstrut&$(1.77\pm 0.07)\times 10^{-3}$&$(3.83\pm 0.11)\times 10^{-2}$&$(4.15\pm 0.19) \times 10^{-2}$\\
    \hline
  \end{tabular}
  \caption{\label{tab:1} The lengths of the sides of the unitarity
    triangles, defined in Eqs.~\eqref{eq:3}.}
\end{table}

The triangle inequality
\begin{equation}
  \label{eq:4}
  \abs{l_{2}-l_{3}}\leq l_{1}\leq (l_{2}+l_{3}).
\end{equation}
is the test that the set of the lengths $\{l_{1},l_{2},l_{3}\}$ forms
a triangle. If we check whether the central values of $l_{i}$ in
Table~\ref{tab:1} form a triangle and fulfill the triangle inequality
in Eq.~\eqref{eq:4}, then only the triangles $\Delta_{2}$ and
$\Delta_{5}$ fulfill these conditions. The angles
$\phi_{k}^{\Delta_{i}}$ for these triangles are equal
\begin{equation}
  \begin{aligned}
    &\phi_{1}^{\Delta_{2}}=(24.21\pm 2.15)^{\circ},\;
    \phi_{2}^{\Delta_{2}}=(92.46\pm 10.12)^{\circ},\;
    \phi_{3}^{\Delta_{2}}=(63.32\pm 9.99)^{\circ},\\[4pt]
    &\phi_{1}^{\Delta_{5}}=(63.41\pm 9.12)^{\circ},\;
    \phi_{2}^{\Delta_{5}}=(90.21\pm 9.09)^{\circ},\;
    \phi_{3}^{\Delta_{5}}=(26.37\pm 2.25)^{\circ}.
  \end{aligned}\label{eq:7}
\end{equation}
From Table~\ref{tab:1} and Eqs.~\eqref{eq:7} one can see that the
angles and the side lengths of the triangles~$\Delta_{2}$
and~$\Delta_{5}$ are very close and this could be expected from the
fact that the deviations from the symmetry of the CKM matrix are
small. From~\eqref{eq:14} one knows that unitarity implies that
$\phi_{2}^{\Delta_{2}} = \phi_{2}^{\Delta_{5}}$ and within an error it
is indeed a case.

The fact that the triangle inequality~\eqref{eq:4} is not fulfilled by
the central values $l_{i}$ of the triangles~$\Delta_{1}$,
$\Delta_{3}$, $\Delta_{4}$ and~$\Delta_{6}$ (see also the figures of
the unitarity triangles in~\cite{bib:1b}) cannot be interpreted as a
violation of the unitarity of the CKM matrix, because
Eqs.~\eqref{eq:4} are fulfilled by these triangles within one standard
deviation.

\section{Analysis of the triangles $\Delta_{2}$ and $\Delta_{5}$}
\label{sec:analys-triangle-delt}

Let us now discuss in more detail the unitarity triangles~$\Delta_{2}$
and $\Delta_{5}$. From Section~\ref{sec:unitarity-triangles} we know
that the experimental information about~$\Delta_{2}$ consists of the
lengths~$l_{i}$ of the sides of the triangle, given in
Table~\ref{tab:1} and of the
angles~$\phi_{i}^{\Delta_{2}}=\phi_{i}^{\text{exp}}$, given in
Eq.~\eqref{eq:8}. This means that there are~6 experimental data for~3
degrees of freedom. We will analyze the compatibility of the data by
taking the experimental input for $\phi_{i}^{\text{exp}}$ and
$\Delta\phi_{i}^{\text{exp}}$ from Eq.~\eqref{eq:8} and for
$l_{i}^{\text{exp}}$, $\Delta l_{i}^{\text{exp}}$ from
Table~\ref{tab:1} and minimizing the function $\chi^{2}$
\begin{multline}
  \label{eq:5}
  \chi^{2}=\qty(\frac{\sin(2\phi_{1})-\sin(2\phi_{1}^{\text{exp}})}
  {\Delta\sin(2\phi_{1}^{\text{exp}})})^{2}
  +\qty(\frac{\phi_{2}-\phi_{2}^{\text{exp}}}
  {\Delta\phi_{2}^{\text{exp}}})^{2}
  +\qty(\frac{\phi_{3}-\phi_{3}^{\text{exp}}}
  {\Delta\phi_{3}^{\text{exp}}})^{2}\\
  +\qty(\frac{l_{1}-l_{1}^{\text{exp}}}{\Delta
    l_{1}^{\text{exp}}})^{2}
  +\qty(\frac{l_{2}-l_{2}^{\text{exp}}}{\Delta
    l_{2}^{\text{exp}}})^{2}
  +\qty(\frac{l_{3}-l_{3}^{\text{exp}}}{\Delta
    l_{3}^{\text{exp}}})^{2}
\end{multline}
with respect to the $l_{1}$, $l_{2}$, $l_{3}$. The values of the
fitted parameters are
\begin{equation}
  \label{eq:9}
  \begin{aligned}
    l_{1}&=(3.45 \pm 0.09)\times 10^{-3},\\
    l_{2}&=(8.99 \pm 0.16)\times 10^{-3},\\
    l_{3}&=(8.50 \pm 0.18)\times 10^{-3}
  \end{aligned}
\end{equation}
and the value of the $\chi^{2}$ and the \textit{p}-value of the fit
are
\begin{equation}
  \chi^{2}=2.91,\quad \text{\textit{p}-value} = 0.41\label{eq:10}
\end{equation}
so the experimental data for the triangle $\Delta_{2}$ are compatible
with the assumption that $l_{1}^{\text{exp}}$, $l_{2}^{\text{exp}}$
and $l_{3}^{\text{exp}}$ form a triangle, whose angles are
$\phi_{1}^{\text{exp}}$, $\phi_{2}^{\text{exp}}$
and~$\phi_{3}^{\text{exp}}$.

The triangle $\Delta_{2}$ thus obtained, whose lengths of the sides
$l_{i}$ are given in Eq.~\eqref{eq:9} has the following values of the
angles
\begin{equation}
  \label{eq:15}
  \begin{aligned}
    \phi_{1}^{\Delta_{2}}&=(22.54\pm 0.72)^{\circ} ,\\
    \phi_{2}^{\Delta_{2}}&= (86.77\pm 4.1)^{\circ},\\
    \phi_{3}^{\Delta_{2}}&= (70.69\pm 3.9)^{\circ} .\\
  \end{aligned}
\end{equation}
By comparing the values of the lengths given in Eqs.~\eqref{eq:9} with
those of Table~\ref{tab:1} we note that the errors are significantly
reduced. The error reduction also occurs for the values of angles in
Eqs.~\eqref{eq:15} and~\eqref{eq:7}.

The analysis for the triangle $\Delta_{2}$ cannot be applied for the
remaining triangles, because their angles have not been
measured. Additionally the lengths of the sides of some of the
triangles violate the triangle inequality Eq.~\eqref{eq:4}, so
strictly speaking they do not form the triangles. To improve this
situation we apply two additional constraints in the fit of the
lengths of the sides for those triangles:
\begin{equation}
  \label{eq:18}\left.
\parbox{0.85\linewidth}{
  \begin{enumerate}
  \item The range of the fitted values of the lengths have to fulfill the triangle inequality Eq.~\eqref{eq:4}.
  \item The area of the fitted triangle has to be equal to the area
    for the triangle $\Delta_{2}$, which is equal
    \begin{equation*}
      \label{eq:11}
      \text{Area of the triangle $\Delta_{2}$} = (1.464 \pm 0.048)\times10^{-5}.
    \end{equation*}
  \end{enumerate}}\quad\right\}
\end{equation}
Such a fit gives the following result for the lengths of the sides
of triangle~$\Delta_{5}$
\begin{equation}
  \label{eq:12}
  \begin{aligned}
    l_{1}&=(7.75^{+0.18}_{-0.17})\times 10^{-3},\\
    l_{2}&=(8.71 \pm 0.25)\times 10^{-3},\\
    l_{3}&=(3.80^{+0.12}_{-0.10})\times 10^{-3}
  \end{aligned}
\end{equation}
and the value of the $\chi^{2}$ and the \textit{p}-value of the fit
are
\begin{equation}
  \label{eq:16}
  \chi^{2}=0.112,\quad\text{\textit{p}-value}=0.262.
\end{equation}
From the values in Eq.~\eqref{eq:12} we obtain the angles of the
triangle $\Delta_{5}$
\begin{equation}
  \label{eq:13}
  \phi_{1}^{\Delta_{5}}=(62.71\pm 4.4)^{\circ} ,\quad
  \phi_{2}^{\Delta_{5}}= (91.46\pm 5.2)^{\circ},\quad
  \phi_{3}^{\Delta_{5}}= (25.82\pm 1.2)^{\circ} .
\end{equation}
Eqs.~\eqref{eq:15} and~\eqref{eq:13} demonstrate that the fitted
values of the angles are compatible with the relation
$\phi_{2}^{\Delta_{2}}=\phi_{2}^{\Delta_{5}}$ in Eq.~\eqref{eq:14} and
the errors are reduced.

\section{Analysis of the remaining triangles }
\label{sec:analysis-remaining-triangles}

\paragraph{The lengths of the sides\\}
For the triangles $\Delta_{1}$, $\Delta_{3}$, $\Delta_{4}$, and
$\Delta_{6}$ we know only the lengths of the sides, given in
Table~\ref{tab:1} and the angles have not been measured, so the
situation is similar as in the case of the triangle $\Delta_{5}$, so
we use the additional constraints described in Eq.~\eqref{eq:18} for
the determination of the lengths $l_{i}$. The results of the fits are
given in Tables~\ref{tab:2} and~\ref{tab:3}.
\begin{table}[H]
  \centering
  \begin{tabular}[c]{l|l}
    \hline
    \multicolumn{1}{c|}{\Tstrut\Bstrut Triangle $\Delta_{1}$}
    &\multicolumn{1}{c}{Triangle $\Delta_{3}$}\\
    \hline
    \Tstrutt$l_{1}=0.21859 \pm (2.5\times10^{-6})$
    &$l_{1}=(8.55 \pm 0.34)\times10^{-4}$\\
    $l_{2}=0.21831 \pm (2.5\times10^{-6})$
    &$l_{2}=(3.946 \pm 0.009)\times10^{-2}$\\
    \Bstrut$l_{3}=(3.104 \pm 0.023)\times10^{-4}$
    &$l_{3}=(3.988 \pm 0.009)\times10^{-2}$\\
    \hline
    \multicolumn{1}{c|}{\Tstrut\Bstrut Triangle
    $\Delta_{4}$}&\multicolumn{1}{c}{Triangle $\Delta_{6}$}\\
    \hline\Tstrut$l_{1}=0.21967^{+0.000022}_{-0.00018}$
    &$l_{1}=(1.77\pm 0.01)\times10^{-3}$\\
    \Tstrutt\Bstrutt$l_{2}=0.21959^{+0.00018}_{-0.000022} $
    &$l_{2}=(3.875 \pm 0.002)\times10^{-2}$\\
    \Bstrut$l_{3}=(1.565 \pm 0.85)\times10^{-5}$
    &$l_{3}=(4.035 \pm 0.002)\times10^{-2}$\\
    \hline
  \end{tabular}
  \caption{\label{tab:2}The results of the fit of the unitarity
    triangles.}
\end{table}
\begin{table}[H]
  \centering
  \begin{tabular}[c]{cll}
    \hline
    \Bstrut\Tstrut Triangle
    &\multicolumn{1}{c}{$\chi^{2}$}
    &\multicolumn{1}{c}{\textit{p}-value}\\
    \hline
    \Tstrut$\Delta_{1}$&0.0016&\hspace*{15pt}0.032\\
    $\Delta_{3}$&0.75&\hspace*{15pt}0.386\\
    $\Delta_{4}$&1.91&\hspace*{15pt}0.166\\
    \Bstrut$\Delta_{6}$&0.54&\hspace*{15pt}0.462\\
    \hline
  \end{tabular}
  \caption{The values of $\chi^{2}$ and the \textit{p}-values of the
    fits for the unitarity triangles.}
  \label{tab:3}
\end{table}

\paragraph{Angles of the triangles\\}
From Table~\ref{tab:2} we can see that the lengths of the sides
$l_{1}$ and $l_{2}$ of the triangles $\{\Delta_{1},\Delta_{4}\}$ are
very close and it is also true for the pair of the sides $l_{2}$ and
$l_{3}$ of the triangles $\{\Delta_{3},\Delta_{6}\}$. What is more
important for each triangle $\Delta_{i}$, $i=1,3,4,6$ one side is much
shorter than two remaining ones.  This causes that the smallest angle
is determined more precisely than the remaining ones and one can see
that the values of those angles in Table~\ref{tab:4} are compatible
with the relations in Eq.~\eqref{eq:14}. The fitted values of the
angles are given in Table~\ref{tab:4}.
\begin{table}[H]
  \centering
  \begin{tabular}[c]{l|l}
    \hline
    \multicolumn{1}{c|}{\Tstrut\Bstrut Triangle $\Delta_{1}$}
    &\multicolumn{1}{c}{Triangle $\Delta_{3}$}\\
    \hline
    \Tstrutt$\phi_{1}=(154.3 \pm 1.8)^{\circ}$
    &$\phi_{1}=(1.07 \pm 0.12 )^{\circ}$\\
    $\phi_{2}=(25.7 \pm 1.8)^{\circ}$
    &$\phi_{2}=(59.9 \pm 10.1)^{\circ}$\\
    \Bstrut$\phi_{3}=(0.0352 \pm 0.0024)^{\circ}$
    &$\phi_{3}=(119.1 \pm 10.1)^{\circ}$\\
    \hline
    \multicolumn{1}{c|}{\Tstrut\Bstrut Triangle
    $\Delta_{4}$}&\multicolumn{1}{c}{Triangle $\Delta_{6}$}\\
    \hline\Tstrut$\phi_{1}=(120.9 \pm 62.2)^{\circ}$
    &$\phi_{1}=(1.03 \pm 1.02)^{\circ}$\\
    \Tstrutt\Bstrutt$\phi_{2}=(59.1 \pm 62.2)^{\circ} $
    &$\phi_{2}=(23.1 \pm 24.1)^{\circ}$\\
    \Bstrut$\phi_{3}=(0.035 \pm 0.022)^{\circ}$
    &$\phi_{3}=(155.9 \pm 25.1)^{\circ}$\\
    \hline
  \end{tabular}
  \caption{\label{tab:4}The angles of the unitarity triangles,
    obtained from the fitted values of the sides of the triangles.}
\end{table}

\section{Summary and conclusions}
\label{sec:disc-concl}

We have examined all the unitarity triangles of the CKM matrix and
analyzed their properties. Our analysis is independent of the
parameterization of the CKM matrix and it is based on the properties
of the unitarity triangles that follow from the~SM.

There are two types of the SM~predictions concerning the unitarity
triangles:
\begin{enumerate}
\item The unitarity of the CKM~matrix implies
  \begin{equation}
    \label{eq:20}
    \phi_{1}^{\Delta_{i}}+ \phi_{2}^{\Delta_{i}}+ \phi_{3}^{\Delta_{i}}= 180^{\circ}
  \end{equation}
  and it it involves the angles of only \emph{one} unitarity triangle.
\item The other type of the relations, given in Eqs.~\eqref{eq:14},
  involve the angles of \emph{different} unitarity triangles.
\end{enumerate}
Relation of Type~1 given by Eq.~\eqref{eq:20} is a test of the
unitarity of the CKM~matrix and it is tested for one unitarity
triangle and is well satisfied by the experimental data~\cite{bib:1} .

Relations of Type~2 given in Eqs.~\eqref{eq:14} follow from the
structure of the~SM and they do not depend on the specific properties
of the CKM matrix, like unitarity. If the CKM matrix exists, then
Eqs.~\eqref{eq:14} have to be fulfilled. On the other hand, if any of
the relations in Eqs.~\eqref{eq:14} are not experimentally fulfilled,
then the description of of the flavour-changing weak interaction by
the CKM matrix is invalid!  It should be stressed that
relations~\eqref{eq:14} constitute a very powerful test of the~SM. If
any pair of angles in those relations is measured and they are not
equal, then the whole SM has to be revised.

Our derivation of the values of the angles of the unitarity triangles
assumed the existence of the CKM matrix so the consistency of our
results for the unitarity angles in Eqs.~\eqref{eq:15}, \eqref{eq:13}
and Table~\ref{tab:4} is just the consistency of our procedure and
calculations.

An important element of our procedure, which we call the
\textit{Method~A} was the assumption that the area of all unitarity
triangles is the same. This fact follows from the unitarity of the CKM
matrix. The experimental input that we use consists of the well
established absolute values of the CKM matrix elements and the angles
of the $\Delta_{2}$ triangle. This information and the equality of the
area of all unitarity triangles allowed us to determine \textit{the
  lengths of the sides and the angles of all the unitarity triangles}.

The alternative way, which we call the \textit{Method~B} of the
determination of all the unitarity triangles is to use the values of
the fitted parameters of the CKM matrix~\cite{bib:1}
\begin{equation}
  \label{eq:19}
  \begin{aligned}
    \sin\theta_{12}&=0.22650 \pm 0.00048,
    \quad&\sin\theta_{13}&=0.00361^{+0.00011}_{-0.00009},\\
    \sin\theta_{23}&=0.04053^{+0.00083}_{-0.00061},
    &\delta&=1.196^{+0.045}_{-0.043}.
  \end{aligned}
\end{equation}
and then to directly obtain the angles of the unitarity triangles
from the standard parameterization~\cite{bib:1,bib:1s} of the CKM
matrix.
\begin{table}[H]
  \centering
  \begin{tabular}[c]{c|c|l|l}
    \hline
    \multicolumn{1}{l|}{\Tstrut\Bstrut Triangle}&angle
    &\multicolumn{1}{c|}{Method A}
    &\multicolumn{1}{c}{Method B}\\
    \hline
     &\Tstrut$\phi_{1}$&$(154.3\pm1.8)^{\circ}$&--\\
    $\Delta_{1}$&$\phi_{2}$&$(25.7\pm1.8)^{\circ}$&--\\
     &\Bstrut$\phi_{3}$&$(0.0352\pm0.0024)^{\circ}$&--\\
    \hline
     &\Tstrut$\phi_{1}$&$(22.54\pm0.72)^{\circ}$&$(22.53\pm2.55)^{\circ}$\\
    $\Delta_{2}$&$\phi_{2}$&$(86.77\pm4.1)^{\circ}$&$(88.98\pm5.13)^{\circ}$\\
     &\Bstrut$\phi_{3}$&$(70.69\pm3.9)^{\circ}$&$(68.49\pm4.31)^{\circ}$\\
    \hline
     &\Tstrut$\phi_{1}$&$(1.07\pm0.12)^{\circ}$&--\\
    $\Delta_{3}$&$\phi_{2}$&$(59.9\pm 10.1)^{\circ}$&$(67.43\pm44.7)^{\circ}$\\
     &\Bstrut$\phi_{3}$&$(119.1\pm 10.1)^{\circ}$&$(111.47\pm47.2)^{\circ}$\\
    \hline
     &\Tstrut$\phi_{1}$&$(120.9\pm 62.2)^{\circ}$&--\\
    $\Delta_{4}$&$\phi_{2}$&--&--\\
     &\Bstrut$\phi_{3}$&$(0.035\pm0.022)^{\circ}$&--\\
    \hline
     &\Tstrut$\phi_{1}$&$(62.71\pm4.4)^{\circ}$&$(67.43\pm1.56)^{\circ}$\\
    $\Delta_{5}$&$\phi_{2}$&$(91.46\pm5.2)^{\circ}$&$(88.98\pm35.8)^{\circ}$\\
     &\Bstrut$\phi_{3}$&$(25.82\pm1.2)^{\circ}$&--\\
    \hline
     &\Tstrut$\phi_{1}$&$(1.04\pm0.81)^{\circ}$&--\\
    $\Delta_{6}$&$\phi_{2}$&$(23.5\pm19.2)^{\circ}$&--\\
     &\Bstrut$\phi_{3}$&$(155.5\pm20.0)^{\circ}$&--\\
    \hline
  \end{tabular}
  \caption{Comparison of the angles of the unitarity triangles
    obtained by two methods. \textit{Method~A} is the one used in this
    paper and the \textit{Method~B} is the direct calculation from the
    standard parameterization of the CKM matrix. The entries ``--''
    mean that the calculated error was larger than the calculated
    value.\label{tab:5}}
\end{table}
In Table~\ref{tab:5} we compare the angles of the unitarity triangles,
which were obtained with our method (\textit{Method~A}) and by the
direct approach (\textit{Method~B}) and we see that the
\textit{Method~B} produces very limited results: one can fully
determine only one unitarity triangle and two angles of the triangles
$\Delta_{3}$ and~$\Delta_{5}$. It should be noted that the predictions
of both methods coincide (within one standard deviation) for the
available angles. The experimental values for the angles of the
$\Delta_{2}$ triangle also are equal (within one standard deviation)
to the predictions of those angles. The fit of the triangle
$\Delta_{4}$, with $\chi^{2}=1.91$ presents some tension with the
existing experimental data.

The progress in the experimental situation and a possibility of a
measurement of the triangle $\Delta_{5}$ can be brought in the
$B$-meson experiments~\cite{bib:27a} at KEK with the Belle~II
detector~\cite{bib:1c} or at CERN with the LHCb
detector~\cite{bib:1d,bib:1e}. Also the theoretical analysis
like~\cite{bib:1f} on removing tensions present in the present data
may bring an important advance in our knowledge of the CKM~matrix.

Any deviation from the predicted values of the angles would be a sign
of the violation of the CKM matrix unitarity and a confirmation of the
$3\sigma$ tension~\cite{bib:1} for the unitarity prediction for the
first row of the CKM matrix and any violation of the relations in
Eq.~\eqref{eq:14} would be a sign of a serious contradiction with
predictions of the~SM.

\section*{Acknowledgment}
\label{sec:acknowledgement}
Supported in part by Proyecto SIP: 20221030, Secretaría de
Investigación y Posgrado, Beca EDI y Comisión de Operación
y Fomento de Actividades Académicas (COFAA) del Instituto
Politécnico Nacional (IPN), México.

\end{document}